# Multi-Level Languages are Generalized Arrows [*]

Adam Megacz

UC Berkeley

megacz@berkeley.edu

## Abstract

Multi-level languages and `Arrows` both facilitate metaprogramming, the act of writing a program which generates a program. The `arr` function required of all `Arrows` turns arbitrary host language expressions into guest language expressions; because of this, `Arrows` may be used for metaprogramming only when the guest language is a superset of the host language. This restriction is also present in multi-level languages which offer unlimited cross-level persistence. The converse restriction, that the host language is a subset of the guest language, is imposed for multi-*stage* languages – those multi-level languages with a `run` construct.

This paper introduces *generalized arrows* and proves that they generalize `Arrows` in the following sense: every `Arrow` in a programming language arises from a generalized arrow with that language's term category as its codomain (Theorem 4.5.4). Generalized arrows impose no containment relationship between the guest language and host language; they facilitate *heterogeneous* metaprogramming. The category having all generalized arrows as its morphisms and the category having all multi-level languages as its morphisms are isomorphic categories (Theorem 4.7.8). This is proven formally in Coq, and the proof is offered as justification for the assertion that *multi-level languages are generalized arrows*.

Combined with the existence of a particular kind of retraction (Definition 4.8.3) in the host language, this proof can be used to define an invertible translation from two-level terms to one-level terms parameterized by a generalized arrow instance. This is ergonomically significant: it lets guest language *providers* write generalized arrow instances while the *users* of those guest languages write multi-level terms. This is beneficial because implementing a generalized arrow instance is easier than modifying a compiler, whereas writing two-level terms is easier than manipulating generalized arrow terms.

Haskell is one example of a host language with the necessary kind of retraction. A modified version of GHC with multi-level terms is offered[1] as a proof of concept; the Haskell extraction of the Coq proofs mentioned above has been compiled into this modified GHC, and is made available as a new flattening pass.

## 1. Introduction

The next section (Section 2) surveys related work. The following three sections cover the same topics three times, each time from a different perspective:

- Section 3 reviews `Arrows`, introduces generalized arrows, reviews multi-level languages, and previews the translation from the latter to the former. This is done by example, in a completely informal way, giving an idea of the programmer's perspective and some sample code. Section 3 contains no theorems, proofs, or formal definitions and is not rigorous in any sense.

- Section 4 defines generalized arrows, multi-level languages, and their respective categories, and proves that these categories are isomorphic. All definitions and theorems in this section have been formalized in Coq.

- Section 5 describes a proof-of-concept implementation, instantiating the theorems of the previous section for an extension of System FC [SCJD07, post-publication Appendix C] and using the Haskell extraction of the resulting Coq proofs as part of a modified GHC. All of the examples from Section 3 are included as examples with the compiler.

Section 6 summarizes the paper's results, highlights known deficiencies, and mentions a few directions for future work.

## 2. Related Work

One of the earliest examples of language support for metaprogramming is the `eval` primitive, present from the earliest days of LISP. However, the `eval` primitive lacked access to the state of the current process; Smith [Smi83] remedied this, but did so via a definition which was self-referential in the sense that it used reflection to

```
pow n = if n==0
        then <[ \x -> 1 ]>
        else <[ \x -> x * ~~(pow (n - 1)) x ]>
```

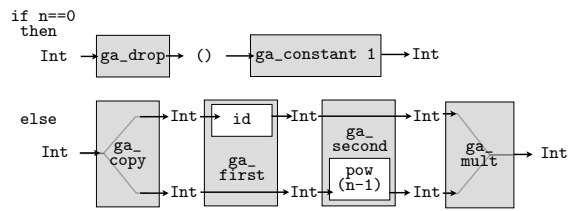

```
pow' n = if n==0
         then    ga_drop                >>>
                 ga_constant 1
         else    ga_copy                >>>
                 ga_first  id           >>>
                 ga_second (pow' (n-1)) >>>
                 ga_mult
```

**Figure 1.** An example two-level program `pow`, which computes $x^n$ (top), its flattened representation as a generalized arrow term (bottom), and a diagrammatic visualization of the terms produced by each of the two branches of the `if` clause (center). Identifiers prefixed with "`ga_`" come from type classes described in Section 3.3. The intuition behind the translation is explained in Figure 5, and Section 4 specifies the translation in detail.

---

[*] This work was supported by a National Science Foundation graduate fellowship, Sun Microsystems Labs, and Oracle Corporation.

[1] http://www.cs.berkeley.edu/~megacz/garrows/
http://www.cs.berkeley.edu/~megacz/coq-categories/

explain reflection. Wand later [Wan86] provided a non-circular explanation of this phenomenon.

More recent research on metaprogramming, specifically multi-level programming in a typed setting, can be divided into *intensional* metaprogramming, which permits inspection of terms from a later level, and *non-intensional* metaprogramming, which does not. Although the former offers additional capabilities for decomposing code which has already been assembled, it requires sophisticated mechanisms for maintaining hygiene and ensuring that $\alpha$-equivalent terms are truly indistinguishable. Most work in this area has dealt with languages that have an intrinsic notion of freshness and name exchange based on nominal logics, as in FreshML [SPG03]. The present paper deals only with non-intensional metaprogramming. Although non-intensional metaprogramming lacks the ability to inspect code terms, it does have a counterbalancing advantage: the compiled implementation is free to perform "reduction under lambda" on later-stage terms, altering their syntactic structure. Because this syntactic structure cannot be examined by earlier stages, such reductions are safe. Research in this area has clustered mainly around two approaches: level annotations and modal types.

Level annotations were originally motivated by the annotations used for manual binding-time analysis; these annotations and their use for explicit staging were first codified in MetaML [TS00] which was developed in quite a large number of subsequent papers, culminating in a full-featured implementation called MetaOCaml [CLG$^+$01]. A major advance was the introduction of *environment classifiers* by Taha and Nielsen [TN03]. By careful use of these classifiers – type variables which are never instantiated except with other type variables – one can express, in the type system, the fact that a given term from a later stage has no free variables. This, in turn, can be used to ensure at compile time that the `run` operation is used safely. More recently, work has been done on inference for these classifiers [CMT04].

Modal types served as an early foundation for investigations into multi-level programming, beginning with the use of temporal logics by Davies and Pfenning [Dav96], who note that "Moggi's computational $\lambda$-calculus only distinguishes values from computations and does not allow us to express stage separation." Along with Wickline and Lee [WLPD98], they achieved an implementation capable of runtime code generation [WLP98] and identified inlining as an operation which cannot be expressed via S4 types. This led to more sophisticated modal operators, such as that of Nanevski and Pfenning's $\nu^{\Box}$ calculus [NP05]. Along with the metamathematics of contexts [BBM95], this work suggests a path toward a unified account of environment classifiers as modal types.

Most work on giving category-theoretic semantics to multi-level languages has been based on indexed categories. Jacobs [Jac91] appears to have been the first to advocate their use to represent contexts, creating a separate category for each possible context to the left of the turnstile. Indexed categories were also used by Moggi in the form of presheaves to investigate two-level languages [Mog97]. This work was later extended with Harper and Mitchell to explain the phase distinction in the ML module system [HMM96].

Kameyama, Kiselyov, and Shan [KKcS08] provide one translation from homogeneous multi-stage terms into single-stage terms. The resulting encoding uses cartesian tuples to represent contexts, which is workable so long as no substructural types are present in the language. Their work was particularly notable in that it directly addressed the influence of weak side effects such as nontermination and extrapolated to stronger side effects which might cause scope extrusion (for example, reference cells). More recently Choi, Aktemur, Yi, and Tatsuta have defined an "unstaging" translation [CAYT11] which can statically analyze homogeneous metaprograms.

The development of `Arrows` began with the premonoidal categories of Power and Robinson [PR97], which were introduced in order to represent contexts as objects in a category; the category Types$(\mathcal{L})$ of this paper is based on the same idea. Power and Thielecke later examined the case where the center of such a category admits a strict monoidal identity-on-objects functor from a cartesian category [PT97], later termed a *Freyd Category*. It was later proved that premonoidal categories which admit such a functors are in bijective correspondence with strong $\kappa$-categories [LPT03, Definition 6.7], a refinement of Hasegawa's $\kappa$-categories [Has95]; these are the term categories of lambda calculi *without* first-class functions. `Arrows` first appeared in functional programming in the work of Hughes [Hug00]. Paterson's paper on `Arrow` notation discusses $\kappa$-categories [Pat01, Section 3.3] and suggests that his "command sublanguage could be viewed as a language for such indexed categories."

Heunen and Jacobs [HJ06] along with Hasuo [JHI09] gave `Arrows` a definition in terms of large categories, viewing them as a monoid in the category of bifunctors **Cat**[$\mathbb{C}\times\mathbb{C}^{\mathrm{op}}, \mathbb{V}$] where $\mathbb{C}$ is $\mathbb{V}$-enriched (along with a "strength"). More recently, Atkey [Atk08] has examined the important role of enrichment in the definition and use of `Arrows`.

## 3. Introduction By Example

### 3.1 `Arrows` in Haskell

From a programmer's perspective, an `Arrow` is a type belonging to the type class below; the `id` and `>>>` declarations come from the `Category` superclass:

```
class Category a => Arrow a where
--id    :: a x x
--(>>>) :: a x y -> a y z -> a x z
  arr   :: (x -> y) -> a x y
  first :: a x y -> a (x,z) (y,z)
```

Instances of this class must obey the laws of [Pat01, Figure 1].

### 3.2 BiArrows in Haskell

`BiArrows` are meant to be `Arrows` with a notion of *inversion*. They were introduced in [ASvW$^+$05] and further examined in [JHI09]. A `BiArrow` is an instance of the following class:

```
class Arrow b => BiArrow b where
  biarr :: (x->y) -> (y->x) -> b x y
  inv   :: b x y -> b y x
```

The `BiArrow` class adds a new constructor `biarr`, which is intended to be used in place of `arr`. It takes a pair of functions which are assumed to be mutually inverse. The `inv` function attempts to invert a `BiArrow` value.

Types belonging to the class `BiArrow` consist of operations which *might be* invertible. Some `BiArrow` values are actually not invertible, so the `inv` operation is partial and may fail at runtime. Since the type of `inv` does not include any way of reporting such failures (e.g., an `Either` type or `Error` monad), the behavior in these circumstances is undefined and must appeal to mechanisms outside the semantics of the language (e.g., `Prelude.error`). The type system is not capable of ensuring that "well-typed programs cannot go wrong" [Mil78] in this way[2]. Moreover, nothing stops a program from

---
[2] Note that aside from failures of `inv` there is a second, separate, issue: the user of a `BiArrow` might supply a pair of functions to `biarr` which are not in fact mutually inverse to each other. However, in this case the program will simply compute the wrong result rather than fail via some

```
class Category g => GArrow g (**) u where
--id           :: g x x
--(>>>)        :: g x y -> g y z -> g x z
  ga_first     :: g x y -> g (x ** z) (y ** z)
  ga_second    :: g x y -> g (z ** x) (z ** y)
  ga_cancell   :: g (u**x)          x
  ga_cancelr   :: g    (x**u)       x
  ga_uncancell :: g      x         (u**x)
  ga_uncancelr :: g      x          (x**u)
  ga_assoc     :: g ((x** y)**z) ( x**(y **z))
  ga_unassoc   :: g ( x**(y **z)) ((x** y)**z )

class GArrow g (**) u => GArrowCopy g (**) u where
  ga_copy      :: g x (x**x)

class GArrow g (**) u => GArrowDrop g (**) u where
  ga_drop      :: g x u

class GArrow g (**) u => GArrowSwap g (**) u where
  ga_swap      :: g (x**y) (y**x)
```

**Figure 2.** The definition for the (GArrow) type class and its three most frequently implemented subclasses. The class `Category` comes from the standard `Control.Category` module.

```
instance Arrow a => GArrow a (,) () where
  ga_first     = first
  ga_second    = second
  ga_cancell   = arr (\((),x) -> x)
  ga_cancelr   = arr (\(x,()) -> x)
  ga_uncancell = arr (\x -> ((),x))
  ga_uncancelr = arr (\x -> (x,()))
  ga_assoc     = arr (\((x,y),z) -> (x,(y,z)))
  ga_unassoc   = arr (\(x,(y,z)) -> ((x,y),z))

instance Arrow a => GArrowDrop a (,) () where
  ga_drop      = arr (\x -> ())

instance Arrow a => GArrowCopy a (,) () where
  ga_copy      = arr (\x -> (x,x))

instance Arrow a => GArrowSwap a (,) () where
  ga_swap      = arr (\(x,y) -> (y,x))

instance Arrow a => GArrowConstant a (,) () t t
  where ga_constant x = arr (\() -> x)
```

**Figure 3.** Instance declaration showing that every Arrow is a generalized arrow. This may be combined this with the instance declaration for `Arrow (->)` in `Control.Arrow` to provide an interpretation in Haskell for multi-level terms in languages which are a subset of Haskell.

passing a `BiArrow` and its class dictionary to a function whose type is `Arrow a=>...`. Such a function would have no reason to suspect that using `arr` rather than `biarr` in such a situation is dangerous! Not only can the use of `BiArrows` lead to undefined behavior, this behavior may be triggered by the composition of two programs which would each have been completely acceptable in isolation.

Unfortunately there is no way to fix this within the framework of `Arrows`, because the `Arrow` type class requires that `arr` be defined for arbitrary functions – even those like $\mathtt{fst} = \lambda(x,y).x$ which cannot possibly have an inverse. Not even the most powerful dependent type system can help with this problem: any restriction on the use of `arr` would result in an implementation which was no longer an `Arrow`. Moreover, the `arr` function is tightly woven into the laws which prescribe the behavior of `Arrows`, so solving the problem is not as simple as replacing `arr` with `biarr`. Finally, the desugaring algorithms for `Arrow` notation [Hug04, Section 3.4][Tea09, Section 7.10] and the `Arrow` calculus [LWY10] rely on unrestricted use of `arr`, so eliminating or restricting `arr` would mean forfeiting those comforts as well.

Perhaps we might admit only certain special cases of `arr`. For example, we might require some value which behaves like (`arr` $\lambda x.x$), but not allow arbitrary functions (like `fst`) to be lifted. But what criteria should we use for deciding which special cases to require? It would seem unwise to make a choice based on any one particular application such invertible programming. Generalized arrows are the result of searching for a well-motivated collection of definitions to fill the void left by removing `arr`.

### 3.3 Generalized Arrows in Haskell

Section 4 will define generalized arrows formally, in arbitrary languages. However, before presenting the abstract constructions, we will first survey one specific case: generalized arrows *in Haskell*, as embodied in the `GArrow` class shown in Figure 2. Like the `Arrow` class, the `GArrow` class is a subclass of `Category`: every `GArrow` supports the `id` and `>>>` operations, and obeys the laws (associativity and neutrality) required of any category. The `GArrow` class also has a `ga_first` function, whose type and laws are identical to those of the `Arrow` class[3], although the accompanying "second" function (`ga_second`) does not have a default definition in terms of `first`.

Unlike the `Arrow` class, the `GArrow` class has a second type parameter `**` of kind $\star \to \star \to \star$, called the *tensor* of the generalized arrow. This plays a role for generalized arrows analogous to the role of pairing `(,)` in the `Arrow` class: using it, one may form binary trees whose leaves are types. Conceptually, one such tree is the type of the "input" and the other is the type of the "output." In contrast to the `Arrow` class, generalized arrows do not assume that this type operator is necessarily a cartesian product. Furthermore, allowing some type g to be an instance of `GArrow` via multiple different tensors will clarify the treatment of sums and products in Section 3.6.1. The third type parameter u, of kind $\star$, plays the role of `()` in `Arrows`; it is called the *unit* of the generalized arrow. In the "trees of types" analogy above the unit type is the empty leaf.

Note the conspicuous absence of the `arr` function for lifting arbitrary Haskell functions into the generalized arrow. In its place, we find six functions: `ga_{un}cancel{l,r}` and `ga_{un}assoc`. These are used to re-arrange the inputs and outputs of a generalized arrow: the `ga_{un}cancel{l,r}` functions add and remove superfluous inputs and outputs (which must be of the unit type) whereas the `ga_{un}assoc` functions "re-parenthesize" the list of inputs[4]. Readers familiar with category theory will notice that this definition closely parallels that of *premonoidal categories* [PR97].

The `GArrow` class has three subclasses which are used quite frequently. `GArrowCopy` allows one to duplicate inputs, `GArrowSwap`

---

extra-semantical error reporting mechanism. Preventing this sort of problem is possible only with a type system rich enough to express facts about equivalence of functions, which typically requires fairly powerful dependent types (such as Coq's). Generalized arrows do not address this second issue.

[3] One could argue that `first` belongs in a class `BinoidalCategory`, a superclass of `Arrow`.

[4] Why use binary trees instead of lists? One motivation is the desire to avoid having to prove facts about the associativity of list-concatenation at the level of types, something which cannot be done in the type systems of many languages.

allows one to change the order of inputs, and `GArrowDrop` allows one to delete inputs of *arbitrary* type by using `ga_drop` to turn them into inputs of the unit type which may then be discarded using `ga_cancel{l,r}`. Note that the `Arrow` class does not permit this distinction: one can produce functions like `ga_drop`, `ga_copy`, and `ga_swap` for *any* `Arrow` instance by using the mandatory `arr` function (the definitions appear in Figure 3). Readers familiar with type theory and proof theory will notice that these definitions closely parallel the structural operations of ordered linear logic [Res09].

The three subclasses above frequently occur together; to reduce verbosity, the class `GArrowSTKC` collects them:

```
class (GArrowDrop g (**) u,
       GArrowCopy g (**) u,
       GArrowSwap g (**) u) =>
       GArrowSTKC g (**) u
```

Instances of `GArrowSTKC` provide implementations of guest languages which are supersets of the Simply Typed Kappa Calculus [Has95].

Conceptually, the nine functions `ga_{un}cancel{l,r}`, `ga_{un}assoc`, `ga_drop`, `ga_copy`, and `ga_swap` exist to "put back" what generalized arrows have lost – relative to the `Arrow` class – as a result of eschewing a mandatory `arr`. This can be made slightly more concrete by taking a look at the instance declaration in Figure 3 which gives every `Arrow` instance a canonical `GArrow` instance. This instance, in conjunction with the standard instance `Arrow (->)`, may be used as an evaluation mechanism for guest languages which happen to be a *subset* of Haskell. The insertion of (sometimes quite a large number of) these re-arrangement function invocations is typically done by the compiler (see Remark 5.1.1). Programmers are responsible only for supplying the implementation for these re-arrangement functions.

### 3.4 Example: BiGArrows

We can now define the class `BiGArrow`, which avoids the problems of the `BiArrow` class by using generalized arrows in place of `Arrow`s:

```
class GArrow g (**) u => BiGArrow g (**) u where
  biga_arr :: (x -> y) -> (y -> x) -> g x y
  biga_inv :: g x y -> g y x
```

The tutorial which accompanies the modified compiler includes an example `BiGArrow` instance which also implements class `GArrowDrop` using a *logging translation* [MHT04].

### 3.5 Multi-Level Terms and Types

Haskell `Arrow`s come with a very convenient syntax [Pat01, Figure 7], modeled on the monadic `do`-notation. Further work has been done on a calculus which types terms of the `Arrow`'s internal language directly [LWY10]. Is there an equivalent for generalized arrows?

There is. However, rather than being a custom syntax designed for a particular type class or category, it turns out than an existing syntax – which is in fact a typed calculus – is not only usable, but ideal. This calculus is the term syntax and type system of multi-level languages [NN92] augmented with environment classifiers [TN03]. This connection will be spelled out in formal detail in Section 4; here we give only the most cursory preview of multi-level terms, their types, and the connection.

Multi-level terms involve two additional expression forms: brackets and escape. Brackets, written `<[...]>`, serve to demarcate an expression of the guest language. For example, the definition below

```
newtype Code x y =
  Code { unC :: forall a. <[ x -> y ]>@a }

instance Category Code where
  id    = Code <[ \x -> x ]>
  f . g = Code <[ \x -> ~~(unC f) (~~(unC g) x) ]>

instance GArrow Code (,) () where
  ga_first  f = Code <[ \(x,y) -> (~~(unC f) x,y)]>
  ga_second f = Code <[ \(x,y) -> (x,~~(unC f) y)]>
  ga_cancell  = Code <[ \(_,x) -> x ]>
  ga_cancelr  = Code <[ \(x,_) -> x ]>
  ga_uncancell = Code <[ \x -> ((),x) ]>
  ga_uncancelr = Code <[ \x -> (x,()) ]>
  ga_assoc    = Code <[ \((x,y),z) -> (x,(y,z)) ]>
  ga_unassoc  = Code <[ \(x,(y,z)) -> ((x,y),z) ]>

instance GArrowDrop Code (,) () where
  ga_drop     = Code <[ \_ -> u ]>

instance GArrowCopy Code (,) () where
  ga_copy     = Code <[ \x -> (x,x) ]>

instance GArrowSwap Code (,) () where
  ga_swap     = Code <[ \(x,y) -> (y,x) ]>
```

**Figure 4.** Instance declaration showing that level 1 terms form a generalized arrow (at level 0). The `newtype` declaration is needed as a hint to the compiler's type inference mechanism.

is the host language representation of a guest language's identity function:

```
guest_id = <[ \x -> x ]>
```

The escape form, written `~~e`, serves to insert one guest language expression into another. For example, the definition below is the host language function which composes two guest language functions:

```
guest_comp f g = <[ \x -> ~~f (~~g x) ]>
```

Multi-level languages also introduce an additional type form: code types, written `<[t]>@c` for some type `t` and type variable `c`. Expressions wrapped in code brackets are given a type of this form, and expressions subject to the escape operator must bear a type of this form. This means that `guest_comp` above has the following type:

```
guest_comp :: forall c.
    <[y->z]>@c -> <[x->y]>@c -> <[x->z]>@c
```

Note the additional type variable, `c`. This is an *environment classifier* [TN03]. In homogeneous multi-level programming, environment classifiers are used to ensure that certain terms of code type contain no free variables, and are therefore safe to execute. In *heterogeneous* metaprogramming an execution mechanism may or may not be available; even when it is unavailable these classifiers still serve an important purpose: a single host program might manipulate terms in multiple different guest languages; unification of environment classifier variables ensures that if one guest language term is pasted into another term (via the escape operator), the same guest language implementation (i.e., generalized arrow) is used to realize them.

### 3.6 Multi-Level Terms to Generalized Arrows and Back

As a preview of the connection between generalized arrows and multi-level languages, consider Figure 4, which shows that level-1 terms may be used to implement the `GArrow` class (at level-0). This class instance is not typically used in practice – it is not involved in

```
                       f >>>> g  = <[ \x -> ~~g (~~f   x) ]>          pow' :: forall g (**) u.
                       firstc  f = <[ \(x,y) -> (~~f x,y) ]>              (GArrowConstant g (**) Int Int,
                       secondc f = <[ \(x,y) -> (x,~~f y) ]>               GArrowSTLC g (**),
                       mult      :: <[ (Int,Int) -> Int ]>@a              GArrowMult g (**) Int) =>
pow n =              pow' n =                                             Int -> g Int Int
 if n==0              if n==0                                          pow' n =
 then <[ \x ->        then <[ \x  -> () ]>           >>>>               if n==0
         1 ]>              <[ \() -> 1  ]>                              then    ga_drop                  >>>
 else <[ \x ->        else <[ \x  -> (x,x) ]>        >>>>                       ga_constant 1
         x *              firstc <[ \x -> x ]>      >>>>               else    ga_copy                  >>>
         ~~(pow (n - 1)) x     secondc (pow' (n-1)) >>>>                       ga_first id             >>>
                           mult                                               ga_second (pow' (n-1)) >>>
      ]>                                                                       ga_mult
```

**Figure 5.** The `pow` program from Figure 1, written three different ways. On the left, the original program, with spacing adjusted to imitate the following two programs (sadly the multiplication operator has no postfix form; ideally it should appear on the blank line). In the middle, the program has been rewritten in equivalent *point-free* form. On the right, code expressions have been replaced with identical definitions (modulo the `newtype` required as a hint to the compiler's type inference engine) from the `GArrow Code` instance of Figure 4, maintaining left-to-right order of expressions. Because no actual code terms appear in the third program, it is polymorphic in the `GArrow` instance and can be used with instances other than `instance Code`.

---

the flattening process, nor do users typically rely on it. However, it is valid code (the modified compiler accepts it) and will help illustrate the flattening process. It provides one possible "translation back" from generalized arrows to multi-level terms.

The formal definition of the flattening process is explained in complete detail in Section 4; what follows is simply a sketch. Consider the `pow` program from Figure 1:

```
pow n = if n==0
        then <[ \x -> 1 ]>
        else <[ \x -> x * ~~(pow (n - 1)) x ]>
```

One might imagine rewriting the program above *in point-free form* as shown in the middle column of Figure 5. The rewritten program has a very important property: all expressions inside code brackets happen to occur in the `GArrow Code` instance of Figure 4. Because of this, we can replace these expressions with functions defined in the `GArrow` class; when instantiated with `instance GArrow Code`, the resulting program will be identical (modulo the `newtype` wrapper required as a hint to the compiler's instance inference mechanism) to the one above. This vague process is *not* the actual algorithm used to perform flattening, though it is based on the same idea.

### 3.6.1 Beyond `GArrowSTKC`

Generalized arrow instances which offer "constant values" implement the class `GArrowConstant` below:

```
class GArrow g (**) u =>
      GArrowConstant g (**) u t r where
  ga_constant  :: t -> g u r
```

The `ga_constant` function takes a value of type `t` and yields a generalized arrow whose input is the unit type `u` and whose result is type `r`. This return value is much like a Haskell function of type `()->r`. Note that `t` and `r` are not required to be the same type; a declaration `instance GArrowConstant g (**) u t r` asserts that the generalized arrow uses type `r` to *represent* constants whose type *in Haskell* is `t`. The "internal language" of the generalized arrow need not have all of Haskell's types, so one can expect that there may be some types `g` and `r` for which no `GArrowConstant g _ _ _ r` instance is possible.

The benefits of separating Haskell's `(,)` from the generalized arrow's tensor can be seen in the `GArrowSum` class, which is roughly analogous to `ArrowChoice` [Hug00, 5.1]:

```
class (GArrow    g (**)  u,
       GArrow    g (<+>) v) =>
       GArrowSum g (**)  u v (<+>) where
  ga_merge :: g (x<+>x) x
  ga_never :: g v       x
```

Note that `g` is required to be a generalized arrow *via two separate tensors*: the `**` tensor is used for representing contexts, and the `<+>` tensor used to represent the *sum* of two types. The `ga_first` and `ga_second` functions on the `<+>` tensor give us the `left` and `right` of the `ArrowChoice` class, and the other functions of that class may be defined in terms of them. The `ArrowChoice` class fixes `Either` as the type operator used to represent sums, whereas `GArrowSum` leaves it abstract. For this reason we need `ga_merge`, which is akin to the obvious function of type `Either a a -> a` and `ga_never`, which produces a disjunct that "never" occurs. The left and right injections of the sum are given by:

```
ga_inl = ga_uncancell >>> ga_first  ga_never
ga_inr = ga_uncancelr >>> ga_second ga_never
```

Note how this example does not assume that the units of the two generalized arrow instances are the same. If this were the case, it would mean that the generalized arrow had a function similar to the Haskell function of type $\forall \alpha . () \to \alpha$. In Haskell the only function of this type is $\lambda x . \bot$ [Wad89]; total functional languages [Tur04] such as Coq have no functions of this type at all. Clearly the ability to separate the units of the two instances is important here.

The class for products is defined dually:

```
class (GArrow    g (**)  u,
       GArrow    g (<*>) v) =>
       GArrowProd g (**)  u (<*>) v where
  ga_prod_copy :: g x (x<*>x)
  ga_prod_drop :: g x v
```

There is an important symmetry here: the type of `ga_merge` is dual to `ga_copy` and the type of `ga_never` is dual to that of `ga_drop`. There are category-theoretic reasons for this: the unit of a coproduct is the initial object, while the unit of a product is the terminal object. Products have diagonal morphisms while coproducts have codiagonal morphisms.

In most cases, instances of GArrowProd will use the same type for both units u and v; this is because cartesian products let you "throw away" either coordinate, and languages without substructural types similarly let you "throw away" any element of the context (weakening). One can add GArrow subclasses for type operators simulating the other connectives of linear logic [Pfe02] – additive conjunction, additive disjunction, multiplicative conjunction, multiplicative disjunction – in a similar fashion; in these cases the units must be different. Another important example occurs in asynchronous dataflow, where a *pair of streams* is quite different from a *stream of pairs* – the latter has more stringent synchronization behavior. The inability to make this distinction forced the Fudgets library [CH93] to co-opt the *coproduct* structure of the underlying type system to represent pairs of streams, causing an anomaly that Paterson notes [Pat01, Section 5.1] in the type of the Fudgets loop function.

## 4. Foundations

Having introduced generalized arrows in a cursory informal manner, these ideas will now be made rigorous. This section will make frequent use of finite binary trees whose leaves are either a distinguished "empty" value or else of some particular sort $S$:

$$\mathsf{Tree}(S) \;=\; \langle\rangle \mid \langle S \rangle \mid \langle \mathsf{Tree}(S), \mathsf{Tree}(S) \rangle$$

This section will state results in terms of programming languages in general, though a few very basic assumptions are necessary. When speaking of a programming language $\mathcal{L}$ we will assume that it has types, terms, and variables. We will assume that a context is a finite binary tree of types. We will assume some equivalence relation $\equiv$ on terms of the same type in the same context; typically this is a denotational semantics or contextual equivalence under some operational semantics, but those are not the only possibilities. We will assume some sort of syntactical substitution mechanism (let-binding, lambda-abstraction and application, explicit substitution, or some other mechanism) on terms which is associative up to $\equiv$ and for which the identity terms are left and right neutral. We will assume that the language comes with some collection of inference rules such that for every term $e$ which has type $\tau$ in context $\Gamma$ there is a natural deduction proof with conclusion $\Gamma \vdash \langle\tau\rangle$ and no open hypotheses and, and furthermore that $e$ can be recovered from this proof.

### 4.1 The Category Types$(\mathcal{L})$

For a particular programming language $\mathcal{L}$ (e.g., Haskell), there is a well-established way to produce a category. Following the unfortunate convention that categories are named after their objects (**Cat**, **Grp**, **Sets**) rather than their morphisms, we have:

**Formalized Definition 4.1.1** (TypesL)  Let $\mathcal{L}$ be a programming language. Types$(\mathcal{L})$ is the following category. **Objects**: finite binary trees with a type of $\mathcal{L}$ at each nonempty leaf. **Morphisms** from $A$ to $B$: a binary tree $T$ of the same shape as $B$, where each leaf of $T$ is a term which has the type at the corresponding leaf of $B$ in the context consisting of *all* the leaves of $A$. **Equality** of morphisms: a pair of morphisms $f, g : A \to B$ are equal if their constituent terms are pairwise ($\equiv$)-equivalent.

**Formalized Theorem 4.1.2** (TypesL_PreMonoidal)  Types$(\mathcal{L})$ is a premonoidal category, where the functor $- \ltimes B$ sends an object $A$ to $\langle A, B \rangle$ and a morphism $f : A \to C$ to $\langle f, \mathsf{id}_B \rangle$ and the functor $A \ltimes -$ sends an object $B$ to $\langle A, B \rangle$ and a morphism $f : B \to C$ to $\langle \mathsf{id}_A, f \rangle$.

This category is not *strict* premonoidal. Its strictification [Lan71, Theorem XI.3.1] – a category in which the objects are *lists* of types rather than binary trees – is more common in the literature, but removes some structure we will need later on. Perhaps the earliest example comes from Lawvere's thesis [Law63], later revisited as [Law96], explaining quantifiers as adjunctions, which led to the general concept being called a Lawvere Theory. More verbosely, Types$(\mathcal{L})$ could be called "the non-strict multi-sorted Lawvere Theory of the term algebra of $\mathcal{L}$". In the context of logic, Awodey refers to this as the formula category [Awo06, 9.5]. Crole refers to it as the classifying category of a theory [Cro94, 4.8.4]. Other names include "clone category" [Ber98, Definition 8.9.4] and "category of derived operations of an algebra." The full subcategory of Types$(\mathcal{L})$ consisting of only those objects which have exactly one leaf is often called the *Lindenbaum category* of an algebra (in this case, the term algebra), and in the case of the language Haskell this is the category most frequently referred to as "**Hask**." Types$(\mathcal{L})$ occurs in *non*-strictified form as a special case of far more general work by Blute, Cockett, and Seely [BCS97].

### 4.2 The Category Judgments$(\mathcal{L})$

Having defined the category of types of a language, we now move to a second category: its category of *judgments*. Let a context be a binary tree of types (i.e. an object of Types$(\mathcal{L})$) and let a judgment be a pair of contexts, written $\Gamma_1 \vdash \Gamma_2$.

**Formalized Definition 4.2.1** (JudgmentsL)  Let Judgments$(\mathcal{L})$ be the following category. **Objects**: finite binary trees with a *judgment* at each non-empty leaf. **Morphisms** from $A$ to $B$: a binary tree $T$ of the same shape as $B$, where each leaf of $T$ is a proof having all the leaves of $A$ as its hypotheses and the corresponding leaf of $B$ as its conclusion. **Equality** of morphisms: a pair of morphisms $f, g : A \to B$ are equal if the expressions recovered from the conclusions of their constituent proofs are ($\equiv$)-equivalent.

Observe that we allow a context (tree of types) rather than merely a single type as the succedent of a judgment. In contrast to natural deduction for classical logic, these trees are conjunctive rather than disjunctive: a proof of $\Gamma \vdash \langle \tau_1, \tau_2 \rangle$ establishes the well-typedness of both $\tau_1$ *and* $\tau_2$.

Like Types$(\mathcal{L})$, this category can be found elsewhere in the literature, under many names: its use in the study of programming languages and type systems originated with Cartmell [Car86, Section 13], who calls this the "category of contexts and relations" $R(U)$ for a term category $U$ and explains how to construct it for cases where the base category has all finite limits (though Types(Haskell) does not) and certain projection morphisms exist – roughly equivalent to assuming that the unit object is terminal. Lambek [Lam69, p79] calls this the *deductive system on a category* (in this case, the category Types$(\mathcal{L})$). Szabo [Sza78, 1.1.26,2.5.3] calls it a *sequential category generated by an unlabeled deductive system* (i.e., category of sequents). Szabo's work also demonstrates that Types$(\mathcal{L})$ (which he calls $Fm(X)$) is isomorphic to a subcategory of this category via the covariant hom-functor embedding. This last result is the closest thing to Theorem 4.3.1 we have been able to find in the literature.

**Formalized Definition 4.2.2** (Judgments_cartesian)  Judgments$(\mathcal{L})$ is a cartesian category.

### 4.3 Relating Judgments$(\mathcal{L})$ and Types$(\mathcal{L})$

Of great importance to the rest of the development is the relationship between Types$(\mathcal{L})$ and Judgments$(\mathcal{L})$:

**Formalized Theorem 4.3.1** (TypesEnrichedInJudgments)  Types$(\mathcal{L})$ is enriched [Kel82, 1.3] in Judgments$(\mathcal{L})$.

**Formalized Definition 4.3.2** (Types_first, Types_second)  The functors $- \ltimes A$ and $A \ltimes -$ which give Types$(\mathcal{L})$ its

6                                                                                                                                                                              *2018/5/29*

premonoidal structure are *Judgments($\mathcal{L}$)-enriched* functors [Kel82, 1.10].

The following simple definition will be helpful:

**Formalized Definition 4.3.3** (`Enrichment`) *An* enrichment, written $\mathbb{C} \in \mathbb{V}$, is a monoidal category $\langle \mathbb{V}, \oplus, I_v \rangle$ and a premonoidal category $\langle \mathbb{C}, \ltimes, \rtimes, I \rangle$ such that $\mathbb{C}$ is the underlying category of some $\mathbb{V}$-enriched category, and all of the functors $A \ltimes -$ and $- \rtimes A$ are $\mathbb{V}$-enriched functors.

Enriched categories are defined in only those categories which are monoidal, so the monoidal structure of $\mathbb{V}$ is a necessary part of this definition. Asking that $\mathbb{C}$ be premonoidal and including the "extra structure" which makes it so as part of the enrichment will simplify later definitions. Lastly, recall the definition of the *center* of a premonoidal category and the fact that it is a monoidal category:

**Formalized Definition 4.3.4** (`Center`) For $\mathbb{C}$ a premonoidal category, the *center of* $\mathbb{C}$, written $Z(\mathbb{C})$, is the collection of all central morphisms of $\mathbb{C}$.

**Formalized Lemma 4.3.5** (`CenterMonoidal`) For $\mathbb{C}$ a premonoidal category, $Z(\mathbb{C})$ is a monoidal subcategory of $\mathbb{C}$.

### 4.4 An `Arrow` in $\mathcal{L}$

Within this framework of languages-as-categories, what is an instance of the `Arrow` class?

**Formalized Definition 4.4.1** (`FreydCategory`) A *Freyd Category*[PT97][PT99] is an identity-on-objects strict symmetric monoidal functor $\mathcal{J}:\mathbb{C} \to Z(\mathbb{K})$ where $\mathbb{C}$ is a cartesian category and $\mathbb{K}$ (called the *Kleisli category*) is symmetric premonoidal.

**Formalized Definition 4.4.2** (`ArrowInProgrammingLanguage`) An *Arrow in a programming language* $\mathcal{L}$ with Types($\mathcal{L}$) cartesian is an enrichment $\mathbb{K} \in \mathbb{V}_\mathbb{K}$ with $\mathbb{V}_\mathbb{K} \subseteq$ Types($\mathcal{L}$) and a Freyd Category $\mathcal{J}$:Types($\mathcal{L}$) $\to Z(\mathbb{K})$ [Atk08, Theorem 3.7].

For example, in Haskell any particular `instance Arrow T` determines a full subcategory of Types(Haskell) consisting of types of the form `T X Y` for any `X` and `Y`. The Kleisli category $\mathbb{K}$ of the `Arrow` is enriched in this subcategory of Types(Haskell).

### 4.5 Generalized Arrows

We are now ready to define a generalized arrow in a programming language:

**Formalized Definition 4.5.1** (`GeneralizedArrowInLanguage`) For programming languages $\mathcal{G}$ and $\mathcal{H}$, a generalized arrow from language $\mathcal{G}$ to language $\mathcal{H}$ is a monoidal functor from Judgments($\mathcal{G}$) to $Z(\text{Types}(\mathcal{H}))$.

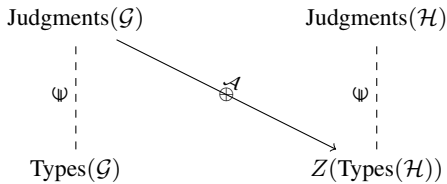

**Remark 4.5.2** Note that the domain of the functor is a judgment category whereas its codomain is [the center of] a *term* category; in this sense, generalized arrows represent the *proof structure of one language within the term structure of another language*.

We can expand the scope of this definition to arbitrary enrichments:

**Formalized Definition 4.5.3** (`GeneralizedArrow`) For enrichments $\mathbb{K} \in \mathbb{V}_\mathbb{K}$ and $\mathbb{C} \in \mathbb{V}$, a generalized arrow from $\mathbb{K} \in \mathbb{V}_\mathbb{K}$ to $\mathbb{C} \in \mathbb{V}$ is a monoidal functor from $\mathbb{V}_\mathbb{K}$ to $Z(\mathbb{C})$.

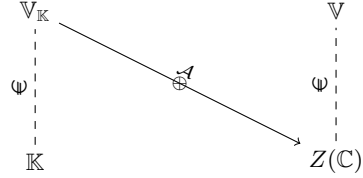

**Formalized Theorem 4.5.4** (`ArrowsAreGeneralizedArrows`) If $\mathcal{J}$:Types($\mathcal{L}$) $\to Z(\mathbb{K})$ is an `Arrow` in programming language $\mathcal{L}$, the inclusion functor from the enriching category of $\mathbb{K}$ to Types($\mathcal{L}$) is a generalized arrow from $Z(\mathbb{K}) \in \mathbb{V}_\mathbb{K}$ to Types($\mathcal{L}$) $\in$ Judgments($\mathcal{L}$):

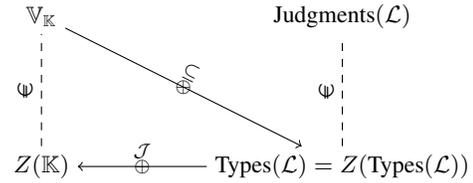

Note that Types($\mathcal{L}$) $= Z(\text{Types}(\mathcal{L}))$ because the domain of a Freyd Category must be cartesian (all morphisms are central).

**Remark 4.5.5** Often, given an `Arrow` $\mathcal{J}$ in a programming language as above, one can cook up a language $\mathcal{L}^A$ such that $Z(\mathbb{K}) \cong \text{Types}(\mathcal{L}^A)$ and $\mathbb{V}_K \cong \text{Judgments}(\mathcal{L}^A)$, making the result of Theorem 4.5.4 not just a generalized arrow between enrichments (Definition 4.5.3) but also a generalized arrow between *languages* $\mathcal{L}^A$ and $\mathcal{L}$ (Definition 4.5.1).

**Remark 4.5.6** A great deal of work has been devoted to isolating the right definition of an `Arrow` in an *arbitrary category* [JHI09, JH06, HJ06, Atk08, Asa10, AH10], such that `Arrows` in programming languages occur as a special case. These `Arrows`-in-arbitrary-categories are outside the scope of the present paper; focus here is limited to those `Arrows` which occur in a programming languages. For example, `Arrows` on the category of sets or on the category of vector spaces are not considered.

### 4.6 Multi-Level Languages

Having defined generalized arrows, we now turn to multi-level languages. First, however, a definition in terms of arbitrary enrichments:

**Formalized Definition 4.6.1** (`Reification`) Given two enrichments $\mathbb{K} \in \mathbb{V}_\mathbb{K}$ and $\mathbb{C} \in \mathbb{V}$, a reification from $\mathbb{K} \in \mathbb{V}_\mathbb{K}$ to $\mathbb{C} \in \mathbb{V}$ is a monoidal functor $\mathcal{R} : \mathbb{V}_\mathbb{K} \to \mathbb{V}$ such that for every $K \in Ob(\mathbb{K})$ there exists a functor $\exists_K : \mathbb{K} \to Z(\mathbb{C})$ making the following diagram commute up to natural isomorphism:

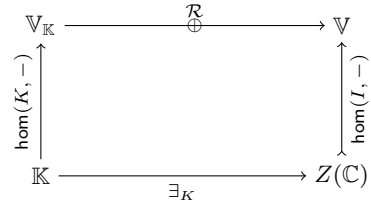

With this definition in place, the definition of a two-level language is straightforward:

**Formalized Definition 4.6.2** (`TwoLevelLanguage`) A two-level language $(\mathcal{G}, \mathcal{H})$ consists of a pair of languages $\mathcal{G}$ and $\mathcal{H}$ and a reification from $\text{Types}(\mathcal{G}) \in \text{Judgments}(\mathcal{G})$ to $\text{Types}(\mathcal{H}) \in \text{Judgments}(\mathcal{H})$.

$$\begin{array}{ccc} \text{Judgments}(\mathcal{G}) & \xrightarrow{\mathcal{R}} & \text{Judgments}(\mathcal{H}) \\ {\scriptstyle\text{hom}(K,-)}\Big\uparrow & & \Big\uparrow{\scriptstyle\text{hom}(I,-)} \\ \text{Types}(\mathcal{G}) & \xrightarrow[\exists_K]{} & Z(\text{Types}(\mathcal{H})) \end{array}$$

**Remark 4.6.3** The essence of a multi-level language is a functor from one Judgments category to another; in this sense, multi-level languages represent the *proof structure of one language within the proof structure of another language*. Compare this with Remark 4.5.2.

The motivation for this definition can be found in literature all the way back to the earliest work on typed multi-level languages. In [NN92, p39] the authors "sketch the form of the general construction [of a two-level language]. For each well-formedness rule or axiom R for types in L, and each binding time $b \in B$, we add the rule or axiom $R^b$;" their specification for $R^b$ – translated into the terminology of category theory – forms the functor $\mathcal{R}$ above when the host language has function types $\Rightarrow$ for which $\text{Hom}(X, Y) \cong \text{Hom}(I, X \Rightarrow Y)$[5].

**Remark 4.6.4** Note that the functors $\mathcal{R}$ and $\exists_K$ are not necessarily full or faithful. This will be revisited in Section 6.

By iterating and taking limits, we arrive at $n$-level languages and $\omega$-level languages:

**Formalized Definition 4.6.5** (`NLevelLanguage`) Every language is a one-level language. An $(n+1)$-level language is a two-level language $(\mathcal{L}_n, \mathcal{L})$ where $\mathcal{L}_n$ is an $n$-level language.

**Formalized Definition 4.6.6** (`OmegaLevelLanguage`) An $\omega$-level language is a function $f$ whose domain is the natural numbers such that $(f(n), f(n+1))$ is a two-level language for every $n \in \mathbb{N}$.

## 4.7 Multi-Level Languages are Generalized Arrows

Although $\text{Types}(\mathcal{L}) \in \text{Judgments}(\mathcal{L})$ gives an enrichment for any programming language, not all enrichments arise this way. The next three properties hold for all enrichments which *do* arise this way (Lemma 4.7.5) from languages with cartesian product types. They capture the formal facts needed to drive the remainder of the proof.

**Formalized Definition 4.7.1** (`MonicEnrichment`) An enrichment $\mathbb{C} \in \mathbb{V}$ is *monic* if there is an object $X$ such that $\text{hom}(X, -) : \mathbb{C} \to \mathbb{V}$ is faithful.

The enrichment of any concrete category in **Sets** is monic. Intuitively, to say that an enrichment is monic is to say that its base category $\mathbb{C}$ is "concrete with respect to" $\mathbb{V}$ (rather than **Sets**). Often $X$ is called a *generator* for $\mathbb{C}$. The enriched category of a monic enrichment is called *well-pointed* if $X$ is a terminal object.

**Formalized Definition 4.7.2** (`SurjectiveEnrichment`) An enrichment $\mathbb{C} \in \mathbb{V}$ is *surjective* if every object of $\mathbb{V}$ is the tensor of finitely many objects each of which is either the unit object $I_v$ of $\mathbb{V}$ or a hom-object of $\mathbb{C}$.

Intuitively, surjective enrichments are those in which the enriching category has no extraneous *objects* unnecessary to the enrichment (note that this says nothing about *morphisms*). This is not a serious limitation; one can simply restrict the focus to the least full subcategory in which $\mathbb{C}$ is enriched.

**Formalized Definition 4.7.3** (`MonoidalEnrichment`) An enrichment $\mathbb{C} \in \mathbb{V}$ is *monoidal* if $\text{hom}(I, -) : Z(\mathbb{C}) \to \mathbb{V}$ is a monoidal functor.

Intuitively, a monoidal enrichment is one in which the base category's tensor is "rich enough" to represent the structure of its own judgments. The most common case is:

**Formalized Theorem 4.7.4** (`CartesianEnrMonoidal`) Any enrichment $\mathbb{C} \in \mathbb{V}$ with both $\mathbb{C}$ and $\mathbb{V}$ cartesian is a monoidal enrichment.

**Formalized Lemma 4.7.5** (`LanguagesWithProductsAreSMME`) If a programming language $\mathcal{L}$ has cartesian contexts (i.e., weakening, exchange, and contraction) (i.e. pairs), then $\text{Types}(\mathcal{L}) \in \text{Judgments}(\mathcal{L})$ is a surjective monic monoidal enrichment.

In a vague sense, surjective monic monoidal enrichments "look enough like programming languages" for the proof to go through.

**Formalized Definition 4.7.6** (`CategoryOfReifications`) The category of reifications is the category with **Objects**: surjective monic monoidal enrichments, **Morphisms** from $E_1$ to $E_2$: reifications from $E_1$ to $E_2$, **Equality** of morphisms: reifications whose $\mathcal{R}$ functors are naturally isomorphic.

**Formalized Definition 4.7.7** (`CategoryOfGeneralizedArrows`) The category of generalized arrows is the category with **Objects**: surjective monic monoidal enrichments, **Morphisms** from $E_1$ to $E_2$: generalized arrows from $E_1$ to $E_2$, **Equality** of morphisms: generalized arrows which are naturally isomorphic as functors. Composition of two generalized arrows $\mathcal{A} : E_2 \to E_3$ and $\mathcal{A}' : E_1 \to E_2$ is given by $\mathcal{A} \circ \text{hom}_{E_2}(I, -) \circ \mathcal{A}'$.

This leads to the main technical result of the paper:

**Formalized Theorem 4.7.8** (`ReificationsAreGArrows`) The category of reifications is isomorphic to the category of generalized arrows.

In intuitive terms, the proof of isomorphism of categories requires a (identity-preserving, composition-respecting) mapping $M_1$ from generalized arrows to reifications, a (identity-preserving, composition-respecting) mapping $M_2$ from reifications to generalized arrows, and proofs that $M_1(M_2(\mathcal{R}))$ is naturally isomorphic to $\mathcal{R}$ and $M_2(M_1(\mathcal{A}))$ is naturally isomorphic to $\mathcal{A}$. Coming up with $M_1$ and $M_2$ is, of course, the hard part. For $M_1$ we simply pre-compose the generalized arrow functor with the $\text{hom}(I, -)$ functor on its codomain. Finding its companion, $M_2$, is more difficult: its construction relies on the fact that, because the domain of the reification functor $\mathcal{R}$ is part of a *surjective* enrichment, we can reason by induction on trees of hom-objects. The base case relies on the fact that the reification functor forms a commuting square with $\text{hom}(I, -)$ (via $\exists_K$) for *every* $\text{hom}(K, -)$, and the inductive step exploits the fact that the codomain enrichment is monoidal, allowing us to "represent" trees of objects. Together these provide an isomorphism between the range of $\mathcal{R}$ as a subcategory and the base category of its codomain; this isomorphism may be post-composed with the reification to produce a generalized arrow. The Coq proof is normative; this is only a sketch.

---

[5] This was the case for all of the systems considered in [NN92], and is also for all typed $\lambda$-calculi. The definition given here – without reference to function types – may be used to define multi-level languages in which the host language does *not* have such function types (for example, languages based on $\kappa$-calculus).

**Definition 4.7.9**   For $\mathcal{R}$ a reification functor, we will write $\hat{\mathcal{R}}$ for $M_2(\mathcal{R})$ and $\hat{\mathcal{A}}$ for $M_1(\mathcal{A}) = \mathsf{hom}(I, -) \circ \mathcal{A}$.

**Remark 4.7.10**   The definitions above require that all objects of the category of generalized arrows and of the category of reifications be surjective monic monoidal enrichments. This is somewhat more restrictive than necessary: the same result holds when the category has as its objects *all* enrichments, but no morphisms $E_1 \to E_2$ unless $E_1$ is surjective and $E_2$ is monic monoidal. This allows for guest languages whose enrichments are not monoidal.

The process of calculating $\hat{\mathcal{R}}$ is loosely analogous to the creation of the instance in Figure 4. Using $\hat{\mathcal{R}}$ to flatten a multi-level term – loosely analogous to the translation from the left column of Figure 5 to the center column – involves walking over a proof in Judgments($\mathcal{H}$) replacing morphisms in the range of $\mathcal{R}$ (i.e. proofs of well-typedness of terms in code-brackets) with the action of $\hat{\mathcal{R}}$ on their preimage under $\mathcal{R}$. The final step in Figure 5 (from the center column to the right hand column) takes advantage of type classes, a feature which is not in every host language. This final step is host-language-specific, and its formal counterpart is the topic of the next section.

## 4.8 Flattening

**Formalized Definition 4.8.1** (`FlatSubCategory`)   Given a two-level language $(\mathcal{G}, \mathcal{H})$ with reification functor $\mathcal{R}$, the full subcategory of Types($\mathcal{H}$) consisting of all objects $X$ such that $\mathsf{Hom}(I, X)$ is *not* in the range of $\mathcal{R}$ is the *flat subcategory* $\mathsf{Flat}(\mathcal{G}, \mathcal{H})$ of the two-level language.

**Formalized Definition 4.8.2** (`RetractionOfCategories`)   A *retraction of categories* $(\mathcal{F}, \mathcal{F}') : A \sqsubseteq B$ is a pair of functors $\mathcal{F} : A \to B$ and $\mathcal{F}' : B \to A$ such that $\mathcal{F}' \circ \mathcal{F}$ is naturally isomorphic to $\mathsf{id}_A : A \to A$.

**Formalized Definition 4.8.3** (`FlatteningOfReification`)   Let $(\mathcal{G}, \mathcal{H})$ be a two-level language with reification functor $\mathcal{R}$. If there exists a retraction $(\mathcal{F}, \mathcal{F}') : \mathsf{Ran}(\hat{\mathcal{R}}) \sqsubseteq \mathsf{Flat}(\mathcal{G}, \mathcal{H})$ from the range of $\hat{\mathcal{R}}$ to the flat subcategory of Types($\mathcal{H}$), then let $\mathcal{F} \circ \hat{\mathcal{R}}$ be the *flattening* of $\mathcal{R}$.

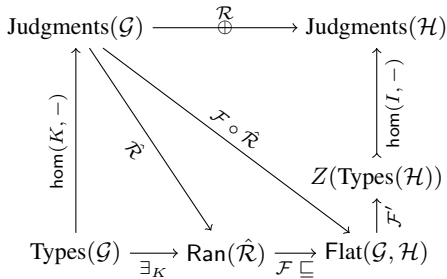

It is immediate that the range of the flattening of a reification falls within the flat subcategory of its codomain, since it arises by post-composing with the retraction. The proof that a retraction $\mathcal{F}$ exists is typically language-specific; the following section addresses the case for Haskell. The flattening process preserves all information present in the original reification in the following sense:

**Formalized Lemma 4.8.4** (`FlatteningIsNotDestructive`)   Post-composing the flattening of a reification with $\mathsf{hom}(I, -) \circ \mathcal{F}^{-1}$ gives back the original reification functor.

## 5. Proof of Concept

To demonstrate the feasibility of these ideas, a modified version of the Glasgow Haskell Compiler (GHC) [Tea09] has been created. This modified GHC offers a new flag, `-XModalTypes`, which adds new expression syntax for code brackets and escape, new type syntax for code types (with environment classifiers), inference of environment classifiers using GHC's existing arbitrary-rank-polymorphism type inference engine [Dim05], and an expansion of the type checker to include the additional typing rules described in this section.

This paper's accompanying Coq script includes a complete formalization of System $\mathsf{FC}^\alpha$ (Section 5.1), a reification from a simple PCF-like language to System $\mathsf{FC}^\alpha$, and a flattening functor for this reification. The Coq code is extracted to Haskell code, which is then compiled in to the compiler, providing a new flag `-fcoqpass` which invokes the flattening transformation. A library `GHC.HetMet.GArrow` is also provided; it includes known-to-the-compiler type classes for generalized arrows.

### 5.1  System $\mathsf{FC}^\alpha$

GHC uses System FC [SCJD07, post-publication Appendix C] as its intermediate language (as of this writing GHC has not yet been updated to use System $\mathsf{FC}_2$ [WVJZ11]). This language is rich enough to capture the type structure of Haskell as well as the large variety of extensions supported by GHC. System $\mathsf{FC}^\alpha$ is little more than the union of the grammars and typing rules of System FC and $\lambda^\alpha$ [TN03]. The remainder of this section explains the modifications made to harmonize the two systems and ensure that its proofs form a category Judgments(System $\mathsf{FC}^\alpha$). System $\mathsf{FC}^\alpha$ is of little direct interest as a type system on its own; it is defined only in order to facilitate the implementation and avoids deviating from the naive union of System FC and $\lambda^\alpha$ except where absolutely necessary. Those deviations are:

1. Inference rules for annotations (Note), recursive let bindings (LetRec), and literals (Lit) have been added because these constructs appear in GHC's `CoreSyn` but not in [SCJD07, post-publication Appendix C].

2. The Alt rule of System FC has been inlined into the Case rule in order to have only a single sort of judgment on expressions.

3. The environments for kinding of type variables ($\Gamma$), classification of coercion variables ($\Delta$), and typing of value variables ($\Sigma$) are kept separate.

4. The branches of a **case** statement, **letrec** bindings, and the environment assigning types to value variables are represented using binary trees and explicit structural manipulations rather than lists and variable names. Because of this, expressions need not appear in judgments – the expression whose well-typedness is witnessed by a proof can be reconstructed, up to $\alpha$-renaming, from the structure of the proof itself. The auxiliary judgment $\Sigma_1 \leadsto \Sigma_2$ asserts that the context $\Sigma_2$ may be produced from $\Sigma_1$ by invocation of structural operations.

5. Rather than represent contexts as type-variable pairs with an auxiliary variable-level map $\sigma$ as in [TN03], contexts are represented as (trees of) type-level pairs with no auxiliary map.

6. Rather than decorate each judgment with a named level, the *succedent* of the judgment is decorated.

**Remark 5.1.1**   The rules for Var and Lit require that the context contain no extraneous entries; this forces the proof-builder to invoke the Weak rule in order to get rid of these entries. Because the flattening functor sends the Weak rule to the `ga_drop` function, this

9                                                                                                                                                              2018/5/29

$$
\begin{aligned}
T &:= \text{type constructors} \\
K &:= \text{data constructors} \\
S_n &:= \text{n-ary type functions} \\
\alpha, \beta &:= \text{type variables} \\
\eta &:= \cdot \mid \alpha, \eta \\
c &:= \text{coercion variables} \\
n &:= \text{notes} \\
p &:= \text{patterns} \\
\kappa, \iota &:= \star \mid \kappa \Rightarrow \kappa \\
\tau, \sigma, \delta &:= \alpha \mid T \mid (\text{->}) \mid (\text{+>}) \\
&\mid \tau\tau \mid \forall \alpha{:}\kappa.\tau \\
&\mid S_n \tau_1 \ldots \tau_n \mid \langle\!\langle\tau\rangle\!\rangle^\alpha \\
\gamma &:= \text{coercions} \\
L_\tau &:= \text{literals of type } \tau \\
x &:= \text{value variables} \\
e &:= x \mid L_\tau \mid \textbf{note } n\ e \\
&\mid \textbf{let } x = e\ \textbf{in } e \\
&\mid \textbf{letrec } \overline{x = e}\ \textbf{in } e \\
&\mid \textbf{cast } \gamma\ e \\
&\mid \textbf{case } e\ \textbf{of } \overline{p \to e} \\
&\mid e\ e \mid \lambda x{:}\tau.e \\
&\mid e\ \tau \mid \Lambda \alpha{:}\kappa.e \\
&\mid e\ \gamma \mid \Lambda^c c{:}\tau{\sim}\tau.e \\
&\mid \langle\!\langle e \rangle\!\rangle \mid {\sim}{\sim}e \\
\Gamma &:= \cdot \mid \alpha{:}\kappa, \Gamma \\
\Delta &:= \cdot \mid c{:}\tau{\sim}\tau, \Delta \\
\Sigma &:= \langle\rangle \mid \langle\Sigma, \Sigma\rangle \mid \langle\tau@\eta\rangle \\
\mathsf{mkTree}(\cdot) &= \langle\rangle \\
\mathsf{mkTree}(x, Y) &= \langle x, \mathsf{mkTree}(Y) \rangle
\end{aligned}
$$

Structural rules:

$$\overline{\langle\langle\rangle, \Sigma\rangle \leadsto \Sigma}\mathsf{CanL} \quad \overline{\langle\Sigma, \langle\rangle\rangle \leadsto \Sigma}\mathsf{CanR} \quad \overline{\Sigma \leadsto \langle\langle\rangle, \Sigma\rangle}\mathsf{uCanL} \quad \overline{\Sigma \leadsto \langle\Sigma, \langle\rangle\rangle}\mathsf{uCanR}$$

$$\overline{\langle\langle\Sigma_1, \Sigma_2\rangle, \Sigma_3\rangle \leadsto \langle\Sigma_1, \langle\Sigma_2, \Sigma_3\rangle\rangle}\mathsf{Assoc} \quad \overline{\langle\Sigma_1, \langle\Sigma_2, \Sigma_3\rangle\rangle \leadsto \langle\langle\Sigma_1, \Sigma_2\rangle, \Sigma_3\rangle}\mathsf{uAssoc}$$

$$\frac{\Sigma_1 \leadsto \Sigma_2}{\langle\Sigma', \Sigma_1\rangle \leadsto \langle\Sigma', \Sigma_2\rangle}\mathsf{Left} \quad \frac{\Sigma_1 \leadsto \Sigma_2}{\langle\Sigma_1, \Sigma'\rangle \leadsto \langle\Sigma_2, \Sigma'\rangle}\mathsf{Right}$$

$$\overline{\langle\Sigma_1, \Sigma_2\rangle \leadsto \langle\Sigma_2, \Sigma_1\rangle}\mathsf{Exch} \quad \overline{\langle\Sigma, \Sigma\rangle \leadsto \Sigma}\mathsf{Cont} \quad \overline{\langle\rangle \leadsto \Sigma}\mathsf{Weak}$$

$$\frac{\Sigma_1 \leadsto \Sigma_2 \quad \Sigma_2 \leadsto \Sigma_3}{\Sigma_1 \leadsto \Sigma_3}\mathsf{Comp} \quad \frac{\Sigma_1 \leadsto \Sigma_2 \quad \Gamma;\Delta;\Sigma_1 \vdash \Sigma}{\Gamma;\Delta;\Sigma_2 \vdash \Sigma}\mathsf{Arrange}$$

$$\frac{\Gamma;\Delta;\Sigma \vdash \langle\tau@\alpha, \eta\rangle}{\Gamma;\Delta;\Sigma \vdash \langle\!\langle\tau\rangle\!\rangle^\alpha@\eta\rangle}\mathsf{Brak} \quad \frac{\Gamma;\Delta;\Sigma \vdash \langle\!\langle\tau\rangle\!\rangle^\alpha@\eta\rangle}{\Gamma;\Delta;\Sigma \vdash \langle\tau@\alpha, \eta\rangle}\mathsf{Esc}$$

$$\frac{\Gamma;\Delta;\Sigma \vdash \langle\tau\rangle}{\Gamma;\Delta;\Sigma \vdash \langle\tau\rangle}\mathsf{Note}(n) \quad \overline{\Gamma;\Delta;\langle\rangle \vdash \langle\tau@\cdot\rangle}\mathsf{Lit}(L_\tau) \quad \overline{\Gamma;\Delta;\langle\tau@\eta\rangle \vdash \langle\tau@\eta\rangle}\mathsf{Var}$$

$$\frac{\Gamma;\Delta;\Sigma_1 \vdash \langle\tau_1@\eta\rangle \quad \Gamma;\Delta;\langle\Sigma_2,\langle\tau_1@\eta\rangle\rangle \vdash \langle\tau_2@\eta\rangle}{\Gamma;\Delta;\langle\Sigma_1,\Sigma_2\rangle \vdash \langle\tau_2@\eta\rangle}\mathsf{Let}$$

$$\frac{\Gamma;\Delta;\Sigma \vdash \langle T\overline\delta@\eta\rangle \quad \theta = [\overline{\alpha \mapsto \delta}] \quad (\forall i)\ K_i : \forall\overline{\alpha{:}\kappa}\forall\overline{\beta{:}\iota}\overline{c{:}\sigma{\sim}\sigma'}.\overline\sigma.\overline{\tau'} \to T\overline\alpha \quad (\forall i)\ \Gamma, \overline{\beta{:}\theta(\iota)}; \overline{\sigma_i{\sim}\sigma'_i}, \Delta; \langle\Sigma, \overline{\theta(\sigma_i)}\rangle \vdash \langle\tau@\eta\rangle}{\Gamma;\Delta;\langle\mathsf{mkTree}(\overline{\Sigma_i}), \Sigma\rangle \vdash \langle\tau@\eta\rangle}\mathsf{Case}$$

$$\frac{\Gamma;\Delta;\Sigma_1 \vdash \Sigma'_1 \quad \Gamma;\Delta;\Sigma_2 \vdash \Sigma'_2}{\Gamma;\Delta;\langle\Sigma_1,\Sigma_2\rangle \vdash \langle\Sigma'_1,\Sigma'_2\rangle}\mathsf{Join} \quad \frac{\langle\tau@\eta'\rangle \in \Sigma_2 \Rightarrow \eta' = \eta \quad \Gamma;\Delta;\langle\Sigma_1,\Sigma_2\rangle \vdash \langle\tau_1@\eta, \Sigma_2\rangle}{\Gamma;\Delta;\Sigma_1 \vdash \langle\tau_1@\eta\rangle}\mathsf{LetRec}$$

$$\frac{\Gamma;\Delta;\langle\Sigma,\langle\tau_x@\eta\rangle\rangle \vdash \langle\tau_e@\eta\rangle}{\Gamma;\Delta;\Sigma \vdash \langle(\text{->})\tau_x\tau_e@\eta\rangle}\mathsf{Abs} \quad \frac{\Gamma;\Delta;\Sigma_2 \vdash \langle(\text{->})\tau_1\tau_2@\eta\rangle \quad \Gamma;\Delta;\Sigma_1 \vdash \langle\tau_1@\eta\rangle}{\Gamma;\Delta;\langle\Sigma_1,\Sigma_2\rangle \vdash \langle\tau_2@\eta\rangle}\mathsf{App}$$

$$\frac{\alpha \notin \Gamma \cup \Delta \cup \Sigma \quad \alpha{:}\kappa, \Gamma;\Delta;\Sigma \vdash \langle\sigma@\cdot\rangle}{\Gamma;\Delta;\Sigma \vdash \langle\forall\alpha{:}\kappa.\sigma@\cdot\rangle}\mathsf{AbsT} \quad \frac{\Gamma \vdash_{\mathsf{TY}} \tau : \kappa \quad \Gamma;\Delta;\Sigma \vdash \langle\forall\alpha{:}\kappa.\sigma@\cdot\rangle}{\Gamma;\Delta;\Sigma \vdash \langle\sigma[\alpha := \tau]@\cdot\rangle}\mathsf{AppT}$$

$$\frac{c \notin \Delta \cup \Sigma \quad \Gamma; c{:}\sigma_1{\sim}\sigma_2, \Delta;\Sigma \vdash \langle\tau@\cdot\rangle}{\Gamma;\Delta;\Sigma \vdash \langle(\text{+>})\sigma_1\sigma_2\tau@\cdot\rangle}\mathsf{AbsC} \quad \frac{\Gamma;\Delta \vdash_{\mathsf{CO}} \gamma : \sigma_1{\sim}\sigma_2 \quad \Gamma;\Delta;\Sigma \vdash \langle(\text{+>})\sigma_1\sigma_2\tau@\cdot\rangle}{\Gamma;\Delta;\Sigma \vdash \langle\tau@\cdot\rangle}\mathsf{AppC} \quad \frac{\Gamma;\Delta \vdash_{\mathsf{CO}} \gamma : \sigma_1{\sim}\sigma_2 \quad \Gamma;\Delta;\Sigma \vdash \langle\sigma_1@\cdot\rangle}{\Gamma;\Delta;\Sigma \vdash \langle\sigma_2@\cdot\rangle}\mathsf{Cast}$$

**Figure 6.** System $\mathsf{FC}^\alpha$, the union of System FC and $\lambda^\alpha$ along with the changes enumerated in Section 5.1. The grammar for coercions $\gamma$ and patterns $p$ and the rules for the type-well-formedness judgment $\vdash_{\mathsf{TY}}$ and coercion-well-formedness judgment $\vdash_{\mathsf{CO}}$ are identical to those of System FC and are not shown; they may be found in [SCJD07, post-publication Appendix C].

requirement is essential to ensuring that the result of the flattening process is well-formed (and well-typed). Similarly, the App, Let, Case, and LetRec rules require that the conclusion context be "partitioned" amongst the hypotheses (Cont/ga_copy) in the correct order (Exch/ga_swap); reconstructing this partitioning from a CoreSyn expression accounts for a significant portion of the complexity in the Coq development, but it is the "engine" of the flattening transformation and it is here that the compiler does the work of piecing together ga_swap, ga_copy, ga_drop, ga_{un}cancel{l,r}, and ga_{un}assoc – the work which makes generalized arrow instances so tedious to use without help from the compiler.

A few conservative restrictions have been imposed in places where there was uncertainty about the interaction between levels and features of System FC; these may be loosened in the future: the variables bound by a **letrec** must have the same named level, literals may appear only at level zero (the compiler comes with special functions to transport literals from level zero to positive levels) and the following expressions may occur only at level zero: the **cast** expression, coercion lambda, coercion application, and **case** expressions which bind coercion variables.

**Formalized Definition 5.1.2** (`SystemFCa`)    System $\mathsf{FC}^\alpha$ is the programming language of Figure 6.

**Formalized Definition 5.1.3** (`PCF`)    PCF is the sublanguage of System $\mathsf{FC}^\alpha$ consisting of proofs which do not use the following rules: Case, Cast, AbsT, AppT, AbsC, AppC. This is roughly equivalent to the *Programming Language for Computable Functions* [Plo77] without integers or booleans.

**Formalized Theorem 5.1.4** (`PCF_SystemFCa_two_level`)    (PCF, System $\mathsf{FC}^\alpha$) is a two-level language for which the the retraction of Definition 4.8.3 exists.

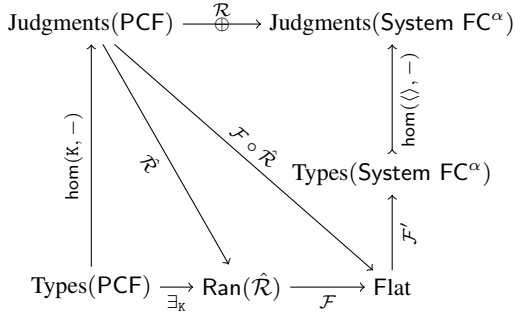

The functor $\exists_K : \text{Types}(\text{PCF}) \to \text{Types}(\text{System FC}^\alpha)$ above sends $f : X \to Y$ in $\text{Types}(\text{PCF})$ to

$\exists_K(f)$ : `forall c. <[K -> X]>@c -> <[K -> Y]>@c`

in $\text{Types}(\text{System FC}^\alpha)$, and the retraction $\mathcal{F}$ sends this to

$\mathcal{F}(\exists_K(f))$ : `forall c. GArrowPCF c => c K X -> c K Y`

## 6. Conclusion and Future Work

Why *heterogeneous* metaprogramming? The effort expended in this paper is largely unnecessary if we are content with guest languages which are the same as – or even supersets of – the host language. The simplest "naturally occurring" example is that of invertible programming, as shown in the running `BiArrow` example of Section 3.2. An invertible programming language cannot include all of Haskell, since Haskell has functions which are quite obviously not invertible (e.g. `fst`).

Beyond applications like `BiArrow` which fit in a paragraph or two lie more appealing examples, including the design of digital circuits (much like Lava [BCSS98, SJRI01, GBK$^+$09]), stream processing of trees [KSK08] based on ordered linear types, and NESL-like data parallelism on platforms *without* global shared memory where (unlike [CLJ07]) parallelized functionals do not have access to the heap (i.e., they are *access-restricted* [Ble95, 10.3]).

### 6.1 Deficiencies

Due to space limitations, the generalized arrow analogs of `ArrowLoop` (`GArrowLoop`), `ArrowApply` (`GArrowApply`, `GArrowCurry`), multi-stage run (`GArrowEval`, `GArrowReflect`), and `arr` (`GArrowReify`) are not described in this paper. They may be found in the modified `base` library, in `GHC/HetMet/GArrow.hs`. The flattening of rule (LetRec) uses the `ga_loopr` and `ga_loopl` definitions of `GArrowLoop`; also used are `ga_curry{l,r}` and `ga_apply{l,r}` from `GArrowCurry` and `GArrowApply`, respectively. `GArrowLoop`, `GArrowCurry`, and `GArrowApply` are all superclasses of `GArrowPCF`.

Although the equivalence of generalized arrows with multi-level languages has been proved in general form as an isomorphism of categories, the flattening lemma relies on the existence of a retraction in the host language. At the moment no necessary-and-sufficient condition is known for the existence of this retraction. This does not impact the soundness of this paper's results, but adds an undesirable extra proof burden to be taken on each time the flattening lemma and corresponding transformation are to be applied to a new pair of guest/host languages. A simple property "P" and proof which works for any pair of guest/host language having property P would be more desirable. The most likely candidate is some variation on "$\mathcal{G}$ is the domain of the initial object of the slice category over $\mathcal{H}$."

Although the current proof scripts establish that there is a reification from PCF to System FC$^\alpha$, the actual functor used in the `-fcoqpass` extraction is constructed directly, rather than by invoking Theorem 4.7.8 and post-composing the resulting functor with a retraction.

### 6.2 Future Work

A number of interesting classes of `Arrows` arise by requiring that the hom-sets of the Kleisli category of the `Arrow` each carry some sort of additional structure. For example, the `ArrowPlus` class assumes that for each pair of types `x` and `y`, if `a` is an instance of `ArrowPlus`, then the values of `a x y` form a monoid, with the `zeroArrow` and `<+>` methods of `ArrowPlus` as the identity element and multiplication operator. `Arrows` with semiring, ring, lattice, boolean algebra, relation algebra, and Kleene algebra structures on their homsets are also possible; the last is particularly useful for concurrency. Even the structure of a category on the homset is possible, making the `Arrow`'s Kleisli category $\mathbb{K}$ a 2-category. All of these possibilities carry over directly to generalized arrows; the question of what the corresponding multi-level language features are has yet to be answered.

The fact that the proofs above work even for reifications whose functors are not full or not faithful (Remark 4.6.4) suggests other interesting directions. A non-faithful reification functor would suggest a guest language with a *finer equivalence relation on terms* than its host language; for example, a guest language with a syntactical notion of equality and a host language in which equality is considered modulo $\alpha$-equivalence. A non-full reification functor suggests a host language which has functions between the images of guest types which, nonetheless, are not themselves the images of guest functions.

In the present implementation it is not possible to determine, solely from the type of a multi-level term, which of `GArrowDrop`, `GArrowSwap`, `GArrowCopy` must be implemented in the generalized arrow supplied to the flattened version of the term. In order to do so, one must track the dereliction, reordering, and duplication of variables in the types of the source language; GHC currently does not have support for this.